\documentclass[12pt]{article}

\usepackage{amssymb}
\usepackage{epsfig}

\def\be{\begin{equation}}
\def\ee{\end{equation}}

\newcommand {\fig} [1] {Fig.~\ref{#1}}

\newcommand {\phiae} {\varphi_{A_{11}}}
\newcommand {\phim} {\varphi_{U_1}}
\newcommand {\phimu} {\varphi_\mu}

\def\lsim{\raise0.3ex\hbox{$\;<$\kern-0.75em\raise-1.1ex\hbox{$\sim\;$}}}
\begin{document}

\title{
\begin{flushright} \small
UWThPh-2003-04\\
HEPHY-PUB 770/2003\\
ZU-TH 06/03\\
hep-ph/0306050  \\
\end{flushright} 
Effect of supersymmetric phases on lepton dipole moments and
rare lepton decays}
\author{A.~Bartl$^1$, W.~Majerotto$^2$, W.~Porod$^3$, D.~Wyler$^3$ \\ \small
       $^1$ Institut f\"ur Theoretische Physik, Universit\"at Wien, \\ \small
         A-1090 Vienna, Austria \\ \small
       $^2$ Institut f\"ur Hochenergiephysik der \"Osterreichischen Akademie
         der Wissenschaften \\ \small
         A-1050 Vienna, Austria\\ \small
       $^3$ Institut f\"ur Theoretische Physik, Universit\"at Z\"urich, \\
         \small
         CH-8057 Z\"urich, Switzerland}
\date{\today}
\maketitle\begin{abstract}
We study the effect of SUSY phases on rare decays of leptons and on
their  magnetic and electric dipole moments. We consider the most general
mass matrices for sleptons within the MSSM including left--right mixing,
flavour mixing and complex phases. We show that the phases also affect
CP even observables. Moreover,
we demonstrate that contrary to common belief the phase of $\mu$
can be large even for slepton masses as small as 200 GeV provided
the lepton flavour violating parameters are complex.
\end{abstract}

\section{Introduction}

The observed neutrino oscillations \cite{skam,sno,kland} are a clear
indication for non-vanishing neutrino masses and violation of
individual lepton numbers. Therefore, one expects flavour violating
effects also for
charged leptons. Furthermore, in analogy to quarks, lepton flavour violation
may also be related to CP violation.

Lepton flavour violation (LFV) in the
charged lepton sector is severely constrained by the stringent experimental
bounds on the branching ratios $BR(\mu \to e \gamma) < 1.2 \cdot 10^{-11}$,
\cite{Brooks:1999pu},
$BR(\tau \to e \gamma) < 2.7 \cdot 10^{-6}$, $BR(\tau \to \mu \gamma) < 1.1
\cdot 10^{-6}$ and rare processes such as $\mu-e$ conversion \cite{pdg}.
The limits on leptonic CP violation, such as the bound of $10^{-27}$
ecm on the electric dipole moment (EDM) of the electron are also quite strong.
However, within a standard model (SM) framework, they are somewhat 
less significant because the leptonic
dipole moments, being   a three--loop effect, are generically small
\cite{shabalin}.

In supersymmetric (SUSY) extensions of the standard model, LFV and
CP violation can also originate in the slepton sector and the
corresponding effects can be generically large. Consequently, rare
processes and $CP$ violation impose significant bounds on the flavour
violating terms in the slepton mass matrices.  The various
phenomenological implications of LFV with {\bf real} mass matrices for
sleptons and sneutrinos were extensively studied ( see
e.~g.~\cite{ref8,ref13,ref14, Porod:2002zy}). The main result is that
despite the stringent experimental bounds on flavour violating lepton
decays, large lepton flavour violating signals are predicted in
production and decays of supersymmetric particles, in particular in
final states containing $e^\pm \tau^\mp$ pairs.  On the other hand,
studies with complex parameters were largely limited to specific SUSY
models, for example the mSUGRA model
\cite{Barbieri:1995tw,strumia,savoy,Branco:2003zy,Masina:2003wt},
or  only some parameters were taken  to be complex \cite{feng,Feng:2003mg}.

In the present paper we study flavour changes and CP violation
in the lepton sector
in the general situation, where all parameters can be complex, 
in particular the LFV entries of the slepton mass matrices. This 
important generalization is quite natural and is motivated by the 
close analogy between quarks and leptons and their
supersymmetric partners. In the $CKM$ matrix the phase is quite large;
the smallness of certain $CP$-violating observables (in the K-system)
is not a result of small phases but of the structure of the theory.

Because CP-violating effects such as electric dipole moments can be
quite large in SUSY, the present experiments impose rather stringent
bounds on phases and it is often suggested that they are small
alltogether \cite{nir}.  Since this view is in a sense contradictory
to the large phase of the standard model, it is desirable to carry out a
general study of flavour and CP violation with complex parameters in order
to see whether the restrictions can be softened.
Furthermore, large leptonic $CP$ violation together with leptogenesis
\cite{lepto} may also be the key to the baryon asymmetry of the
universe. One goal of our work therefore
is to determine whether large phases are
indeed possible and not in contradiction with experiment.
We will demonstrate that in the presence of complex flavour
violating parameters, the usual bounds on the phase of the $\mu$
parameter are no more valid even for slepton masses as
small as 200 GeV.


In general supersymmetric models with soft breaking terms there is a large
number of (complex) parameters. Consequently, each observable can have
contributions from several parameters and no clear statements on their
allowed ranges 
may be possible. As a second goal of our study, we want to show that,
nevertheless, important results can 
be obtained because the present limits on 
rare processes in the lepton sector are so 
strong. Furthermore, several 
experiments with substantially increased sensitivity are planned for the
near future and will lead to even more decisive information.

As there are many parameters involved, we use in this  study 
the first of the so called Snowmass points \cite{Allanach:2002nj} 
as a starting point and  add
flavour violating parameters as well as possible phases. 
%
%
The processes we will study are the 
rare leptonic decays $\mu \to e \gamma$,
$\tau \to e \gamma$ and $\tau \to \mu \gamma$ and the 
electric (EDM)
and magnetic dipole moments (MDM) of $e, \mu$ and $\tau$.
We will use the present experimental bounds,
$d_e < 1.5 \cdot 10^{-27}$ ecm \cite{Commins:gv}, 
$d_\mu < 1.5 \cdot 10^{-18}$ ecm \cite{Bailey:1978mn}, 
$d_\tau < 1.5 \cdot 10^{-16}$ ecm \cite{pdg}.  
For the magnetic moments we assume that the 
supersymmetric contribution is limited by the experimental errors of
$\pm 10^{-12}$ and $\pm 0.058$ for $a_e$ and $a_\tau$ respectively.
For the muon, there are  new
measurements of $a_\mu$ \cite{Sichtermann:2003es}, but there are
still several uncertainties in the theoretical value of  $a_\mu^{SM}$
\cite{jeger,davier}. We will
take the very conservative range $a_\mu^{exp} - a_\mu^{SM} =
43 \cdot 10^{-10}$, which corresponds to the largest deviation in
the calculations.
Future measurements of
$d_e$ \cite{edipol} and $d_\mu$ \cite{mudipol} may 
substantially improve 
the sensitivity to $10^{-29}$ and  $10^{-24}$, respectively. Also 
new experiments for the search of the rare
decay $\mu \to e \gamma$ at the level of $10^{-14}$ \cite{meg}
are underway.

The paper is organized as follows: In the next section we define the
parameters and fix the notation. First we consider in section 3 a
situation without flavour violation but with complex parameters. In
section 4 the general situation with lepton flavour violation and
complex parameters is studied. Some conclusions are drawn in section
5.

\section{The basic parameters}

We assume a general supersymmetric $SU(2) \times U(1) \times SU(3)$ 
model with
the standard soft breaking mass parameters and trilinear scalar couplings 
\cite{Haber:1984rc}.  In the electroweak gaugino sector 
and in the slepton sector the soft breaking part of the Lagrangian
reads as:
\begin{eqnarray}
{\cal L} &=& M^2_{L,ij} {\tilde l}_{L,i}   {\tilde l}^*_{L,j}
         + M^2_{E,ij} {\tilde l}_{R,i}   {\tilde l}^*_{R,j}
         + A_{ij} H_1 {\tilde l}_{L,i} {\tilde l}^*_{R,j}
         + A^*_{ij} H^*_1 {\tilde l}^*_{L,i} {\tilde l}_{R,j} \nonumber \\
  && + M_1 \tilde b \tilde b + M_1^* \bar{\tilde{b}} \bar{\tilde{b}} 
      + M_2 \tilde w^a \tilde w^a + M_2^* \bar{\tilde{w}}^a \bar{\tilde{w}}^a
\label{softy}
\end{eqnarray}
$M_1$ and $M_2$ are the $U(1)$ and $SU(2)$ gaugino mass parameters, 
respectively.
$M^2_{L}$ and $M^2_{E}$ are the soft SUSY breaking mass matrices for
left and right sleptons, respectively, and the $A_{ij}$ are the trilinear soft
SUSY breaking couplings of the sleptons and Higgs boson. 
$M_1$, $M_2$, $M^2_{L,ij} = (M^2_{L,ji})^*$, $M^2_{E,ij} = (M^2_{E,ij})^*$ and 
$A_{ij}$ are complex; note that  $A_{ij} \ne A_{ji}^*$ for $i \ne j$.
The most general charged slepton mass matrix including left-right mixing
as well as flavor mixing is usually written in the form
\begin{equation}
  M^2_{\tilde l} = \left(
    \begin{array}{cc}
      M^2_{LL} &  M^{2\dagger}_{LR} \\
      M^2_{LR} &  M^2_{RR} \\
     \end{array} \right) \, ,
  \label{eq:sleptonmass}
\end{equation}
where the entries are $3 \times 3$ matrices. In terms of the parameters
introduced in (\ref{softy}), they are given by
\begin{eqnarray}
  \label{eq:massLL}
  M^2_{LL,ij} &=& M^2_{L,ij} + \frac{ v^2_d Y^{E*}_{ki} Y^{E}_{kj} }{2}
  + \frac{\left( {g'}^2 -  g^2 \right) (v^2_d - v^2_u) \delta_{ij}}{8}  \, ,\\
  \label{eq:sleptonmassLR}
  M^2_{LR,ij} &=& \frac{ v_d A^*_{ij} - \mu v_u Y^E_{ij} }{\sqrt{2}}  \, ,\\ 
  M^2_{RR,ij} &=& M^2_{E,ij} + \frac{ v^2_d Y^{E}_{ik} Y^{E*}_{jk} }{2}
      -  \frac{ {g'}^2  (v^2_d - v^2_u) \delta_{ij}}{4}  \, .
\end{eqnarray}
The indices $i,j,k=1,2,3$ characterize the flavors $e,\mu,\tau$.
$\mu$ and the $Y^E_{ij}$ are the usual $\mu$ parameter and 
the lepton Yukawa couplings such that $m_l = v_d Y^E_{ll}$. 
$v_u$ and $v_d$ are
the vacuum expectation values of the neutral Higgs fields (with 
$\tan\beta= v_u/v_d$).
In what follows
we will work in a basis where $M_2$ is real and where the lepton Yukawa 
coupling is real and flavour diagonal. Both assumptions can be done
without loss of generality, because (i) only phase differences matter
and (ii) there are no right-handed neutrinos in the low energy spectrum.   
%
 The mass eigenstates $\tilde l_n$ of (\ref{eq:sleptonmass}) are given by 
$\tilde l_n = R^{\tilde l}_{nm} \tilde l'_m$ with 
$l'_m = (\tilde e_L, \tilde \mu_L, \tilde \tau_L,
          \tilde e_R, \tilde \mu_R, \tilde \tau_R)$.

Similarly, one finds for the sneutrinos
\begin{eqnarray}
  M^2_{\tilde \nu,ij} &=&  M^2_{L,ij} 
  + \frac{ \left( g^2 + {g'}^2 \right) (v^2_d - v^2_u) \delta_{ij}}{8}
  \label{eq:sneutrinomass}
\end{eqnarray}
and the corresponding mass eigenstates 
$\tilde \nu_i = R^{\tilde \nu}_{ij}\tilde \nu_j'$ and 
$\tilde \nu_j' = (\tilde \nu_e, \tilde \nu_\mu, \tilde \nu_\tau) $.
We have not explicitly written neutrino Yukawa couplings or (Majorana) masses 
and corresponding soft terms, although neutrino flavour violations were
partly motivating this study. If one assumes the 
sea-saw mechanism for the
(light) left-handed neutrino masses, the right-handed neutrinos and sneutrinos
are very heavy and they practically decouple. The 'left-handed' sneutrino mass
terms will contain contributions from the neutrino Yukawa
couplings, but these will be very tiny playing a role only in the
special case where all soft terms are exactly degenerate; we therefore
neglect them in this work.

The relevant interactions for this study are
\begin{eqnarray}
  \label{eq:CoupChiSfermion}
 {\cal L} &=& \bar l_i ( c^L_{ikm} P_L + c^R_{ikm} P_R)
               \tilde{\chi}^0_k \tilde l_m  
    +  \bar{l_i} (d^L_{ilr} P_L + d^R_{ijr} P_R)
           \tilde{\chi}^-_l \tilde{\nu}_r + h.c.
\label{eq:lag}
\end{eqnarray}
and the couplings $c^L_{ikm}$, $c^R_{ikm}$,
$d^L_{ikm}$, $d^R_{ikm}$ are given
in the Appendix.
\footnote{Similar interactions in the neutrino sector
would give rise to neutrino decays such as $\nu \to \nu' \gamma$ if 
kinematically allowed. Such decays are of interest in cosmology. However, the
estimated lifetimes in our framework are so large that no interesting 
effect is expected.}

\begin{figure}[t]
\setlength{\unitlength}{1mm}
\begin{center}
\begin{picture}(150,30)
\put(-15,-110){\mbox{\epsfig{figure=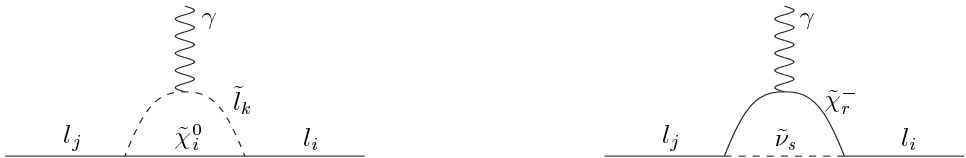,height=24cm,width=19cm}}}
\end{picture}
\end{center}
\caption{Generic diagrams contributing to $\Delta a_l$, $d_l$, 
$l_j \to l_i \, \gamma$.}
\label{fig:diagram}
\end{figure}

The couplings in (\ref{eq:lag}) give, at the 1-loop level, contributions to
the anomalous magnetic moments of the leptons $a_l$, 
the electric dipole moments $d_l$
and to rare lepton decays such as $l_j \to l_i \, \gamma$
if individual lepton number is not conserved.
All these observables are induced by the same amplitude
\begin{eqnarray}
T = i e \epsilon^{\mu *} \frac{q^\nu}{2  m_{l_j}}
    \bar{l}_i \sigma_{\mu \nu} (a^L_{ij} P_L +  a^R_{ij} P_R) l_j
\end{eqnarray}
arising from the diagrams shown in  Fig.~\ref{fig:diagram}. Here we take
$i\le j$. 

The coefficients $a^L_{ij}$ and $a^R_{ij}$ are given by
\begin{eqnarray}
16 \pi^2 a^L_{ij} &=&  \sum_{k=1}^4 \sum_{r=1}^6
  \left( \left( c^L_{ikr}c^{L*}_{jkr}
         \frac{  m^2_{l,j}}{m^2_{\tilde{\chi}^0_k}}
         + c^R_{ikr}c^{R*}_{jkr}
         \frac{  m^2_{l,i}}{m^2_{\tilde{\chi}^0_k}} \right)
         F_1\left(\frac{m^2_{\tilde l_r}}{m^2_{\tilde{\chi}^0_k}} \right)
     +  c^L_{ikr}c^{R*}_{jkr}
         \frac{m_{l,j}}{m_{\tilde{\chi}^0_k}}
         F_3\left(\frac{m^2_{\tilde l_r}}{m^2_{\tilde{\chi}^0_k}} \right)
   \right) \nonumber \\
  &+& \sum_{k=1}^2 \sum_{r=1}^3
  \left( \left( d^L_{ikr} d^{L*}_{jkr}
         \frac{ m^2_{l,j}}{m^2_{\tilde{\chi}^+_k}} 
         + d^R_{ikr} d^{R*}_{jkr}
         \frac{ m^2_{l,i}}{m^2_{\tilde{\chi}^+_k}} \right)
         F_2\left(\frac{m^2_{\tilde \nu_r}}{m^2_{\tilde{\chi}^+_k}} \right)
     +  d^L_{ikr}d^{R*}_{jkr}
         \frac{ m_{l_j}}{m_{\tilde{\chi}^+_k}}
         F_4\left(\frac{m^2_{\tilde \nu_rj}}{m^2_{\tilde{\chi}^+_k}} \right)
   \right) 
\label{eqn:al} \nonumber \\ && \\
a^R_{ij} &=& a^L_{ij}( L \leftrightarrow R)
\label{eqn:ar}
\end{eqnarray}
resulting in the following formulas for $\Delta a_i$, $d_i$ and 
$l_j \to l_i \gamma$:
\begin{eqnarray}
 \Delta a_i &=& \frac{1}{2}
            \mathrm{Re}\left(  a^L_{ii} + a^R_{ii} \right) \\
 \frac{1}{e} d_i &=&
          \frac{1}{2} \mathrm{Im}\left( - a^L_{ii} + a^R_{ii} \right) \\
 \Gamma(l_j \to l_i \gamma) &=& \frac{\alpha m_{l,j}}{16}
    \left( |a^L_{ij}|^2 + |a^R_{ij}|^2 \right)
\end{eqnarray}
where in the last equation we have neglected terms of order
O($m_{l_i}/m_{l_j}$).
The functions $F_i$ and the explicit form of the couplings are listed in
the Appendix.
We consider for consistency only the 1-loop contributions to the EDMs and
the MDMs neglecting the partial 2-loop results for the EDMs presented in
\cite{Chang:1998uc}.

\section{The flavour conserving case}

We begin our investigation with 
the point SPS\#1a \cite{Allanach:2002nj} which
is defined by$M^2_{L,11} = M^2_{L,22} = 202.3^2$~GeV$^2$,
$M^2_{L,33} = 201.5^2$~GeV$^2$, $M^2_{E,11} = M^2_{E,22} = 138.7^2$~GeV$^2$, 
$M^2_{E,33} = 136.3^2$~GeV$^2$,
 $A_{11} = -7.567 \cdot 10^{-3}$~GeV, $A_{22} = -1.565$~GeV,
$A_{33} = -26.326$~GeV, $M_1 = 107.9$~GeV, $M_2 = 208.4$~GeV,
$\mu = 365$~GeV, $\tan\beta = 10$. Note that the $A$ parameters are already
multiplied by the lepton Yukawa couplings.
This gives (for real parameters) the following SUSY contributions to the
observables: $d_e = d_\mu =d_\tau = 0$, $\Delta a_e=6.8 \cdot 10^{-14}$, 
$\Delta a_\mu=2.9  \cdot 10^{-9}$, $\Delta a_\tau= 8.4 \cdot 10^{-7}$.

\begin{figure}
\setlength{\unitlength}{1mm}
\begin{center}
\begin{picture}(150,70)
\put(0,0){\mbox{\epsfig{figure=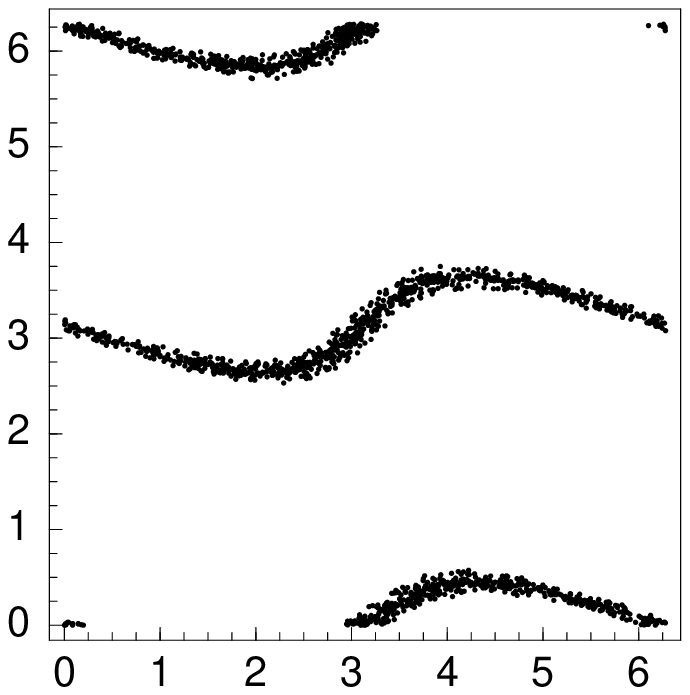,height=7cm,width=7cm}}}
\put(-5,71){\makebox(0,0)[bl]{{\bf a)}}}
\put(0,70){\makebox(0,0)[bl]{$\phimu$}}
\put(70,-3){\makebox(0,0)[br]{$\phim$}}
\put(80,0){\mbox{\epsfig{figure=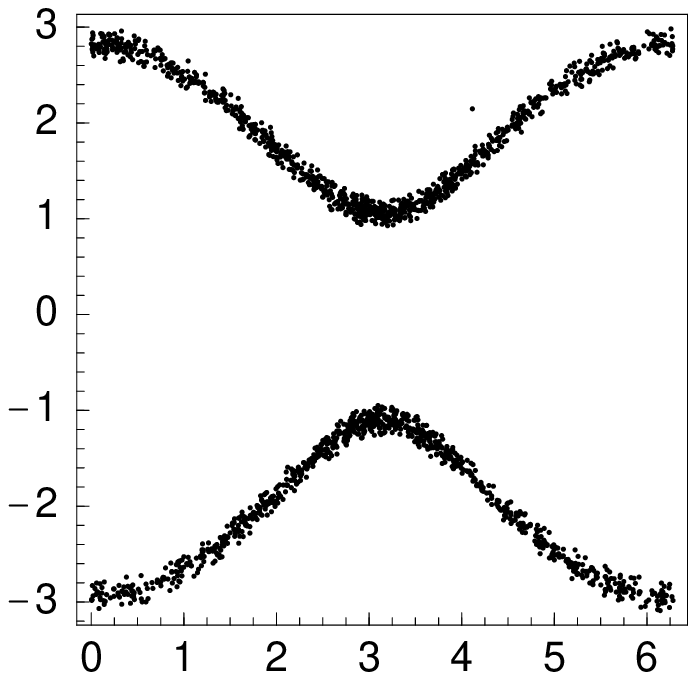,height=7.cm,width=7.cm}}}
\put(75,71){\makebox(0,0)[bl]{b)}}
\put(80,70){\makebox(0,0)[bl]{$10^9 \cdot \Delta a_\mu$}}
\put(150,-3){\makebox(0,0)[br]{$\phim$}}
\end{picture}
\end{center}
\caption{ a) Allowed regions in the $\phimu$--$\phim$ plane
by the electron EDM; b) SUSY contribution
$\Delta a_\mu$ to the anomalous magnetic moment of the muon.}
\label{fig:MuM1plane}
\end{figure}

The flavour conserving
contributions to the leptonic MDM and EDM were extensively discussed in
\cite{amusus1,amusus2,nir,dsus1,dsus2}. 
In the flavour conserving case the phase 
$\phimu$ of the parameter
$\mu$ is severely constrained 
\cite{nir,dsus1,dsus2,Bartl:1999bc,Brhlik:1998zn} by the EDMs of
the electron and  the neutron. There are some
regions in parameter space, where $\phimu$ can be about $\pi/10$ for
slepton masses as light as O(200) GeV if there are 
cancellations between the chargino and neutralino contributions 
\cite{Bartl:1999bc}. In the case of the electron EDM such a cancellation is
due to an interplay of the phases $\phiae$ and $\phim$, where $\phim$
is the phase of the $M_1$ parameter. In \fig{fig:MuM1plane}a we show
the range of the $\phimu$--$\phim$ plane allowed by the electron EDM;
$\phiae$ is varied in the full range. If one fixes this phase, then
the two bands collapse to lines.  Similar results have been found in
\cite{Brhlik:1998zn}. We see how the inclusion of the phase $\phiae$
enlarges the allowed region, but not too much. In
\fig{fig:MuM1plane}b the SUSY contribution $\Delta a_\mu$ to the
anomalous magnetic moment of the muon is shown. The two bands
correspond to the cases where $\phimu$ is centered near $0$ and near
$\pi$. We see that while the EDM leaves a twofold ambiguity for the
phase $\phimu$, the $CP$-conserving anomalous magnetic moment
discriminates between the two values - indeed the lower band is
already excluded and $\phimu$ must be near $\pi$. This has been
observed before \cite{amusus2}. This analysis
shows that phases are also important for CP-conserving observables and
that a combined analysis of all effects is necessary. In case the
theoretical treatment of $a_\mu$ is improved, it might be even possible
to exclude the situation without flavour violation.

Despite the new freedom, $\mu$ is still basically real. Unless the
$A$ parameters are substantially larger, this conclusion remains. However,
bigger values of $|A_{11}|$ are in contradiction with stability arguments for
the potential. As we will see, flavour violating complex
parameters change this picture. 
In addition to SPS\#1a we have considered  also other Snowmass points and
found similar results.

\section{Including flavour violation}

We begin the analysis with a study of pairwise flavour mixings between two  
generations to get an idea which parameters have the largest effect on 
which observables.  The new parameters that need to be introduced 
(say $M^2_{L,12}$, etc.)
are chosen such that the bounds on all observables are 
saturated. 

\subsection{$\tilde e$-$\tilde \mu$  mixing}

Starting from our reference point, we 
add the following flavour violating terms:
$M^2_{L,12} = 0.1$~GeV$^2$, $M^2_{E,12} = 0.1$~GeV$^2$,
$A^l_{12} = 10^{-3}$~GeV, $A^l_{21} =  10^{-3}$~GeV.
With these parameters we get
BR$(\mu \to e \, \gamma) = 1.1 \cdot 10^{-12}$. As mentioned, these
parameters saturate the limits on the branching ratio (we have disregarded
possible subtle cancellations).

\begin{figure}
\setlength{\unitlength}{1mm}
\begin{center}
\begin{picture}(70,70)
\put(0,0){\mbox{\epsfig{figure=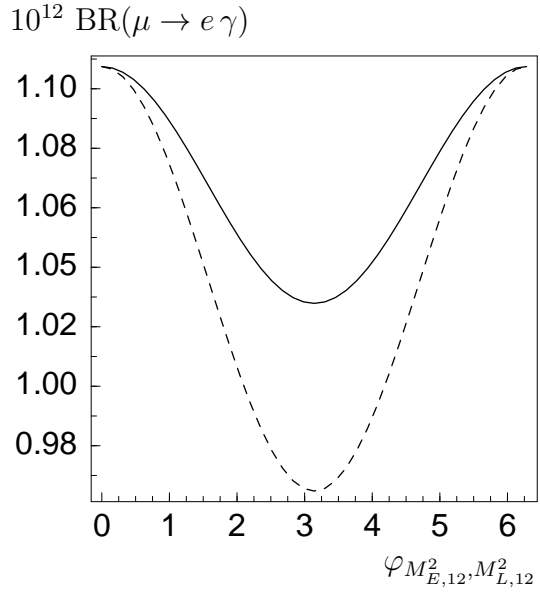,height=7.cm,width=7.cm}}}
\put(0,70){\makebox(0,0)[bl]{$10^{12}$ BR($\mu \to e \, \gamma$)}}
\put(70,-3){\makebox(0,0)[br]{$\varphi_{M^2_{E,12}, M^2_{L,12}}$}}
\end{picture}
\end{center}
\caption{$10^{12}$ 
BR$(\mu \to e \, \gamma)$ as a function of $\varphi_{M^2_{E,12}}$ (full line)
and $\varphi_{M^2_{L,12}}$ (dashed line). The phase $\varphi_\mu$ is set equal to
zero.}
\label{fig:12mixing}
\end{figure}

Due to the relative smallness of the off-diagonal parameters the effect of
the phases is small, as can be seen in \fig{fig:12mixing}.
The effect of the phases of $A^l_{12}$ and $A^l_{21}$ is nearly the
same as that of $\varphi_{M^2_{E,12}}$ and is not shown.
The magnetic and the electric dipole moments are
practically independent of the phases of the flavour violating
parameters and are therefore not explicitly shown.

\subsection{$\tilde e$-$\tilde \tau$  mixing}

\begin{figure}
\setlength{\unitlength}{1mm}
\begin{center}
\begin{picture}(150,70)
\put(0,0){\mbox{\epsfig{figure=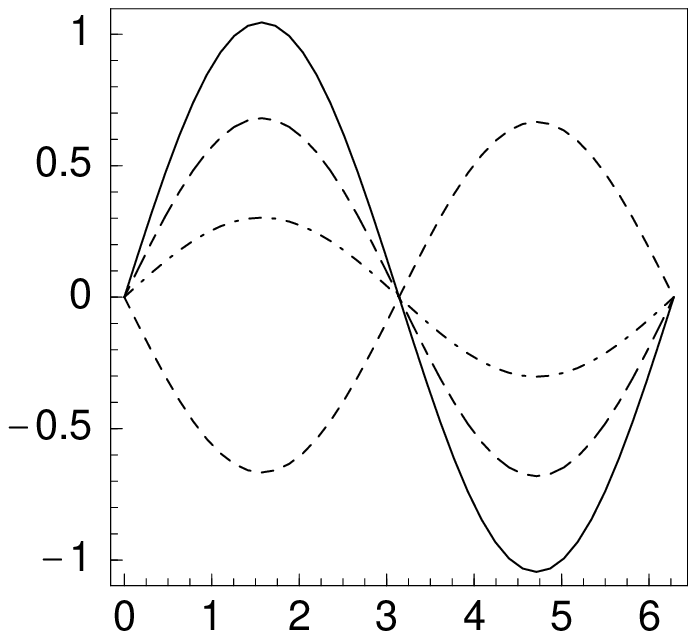,height=7.cm,width=7.cm}}}
\put(-5,71){\makebox(0,0)[bl]{{\bf a)}}}
\put(0,70){\makebox(0,0)[bl]{$10^{23} \cdot d_e$}}
\put(70,-4){\makebox(0,0)[br]{
       $\varphi_{M^2_{E,13}, M^2_{L,13}, A_{13}, A_{31}}$}}
\put(75,71){\makebox(0,0)[bl]{{\bf b)}}}
\put(80,-0){\mbox{\epsfig{figure=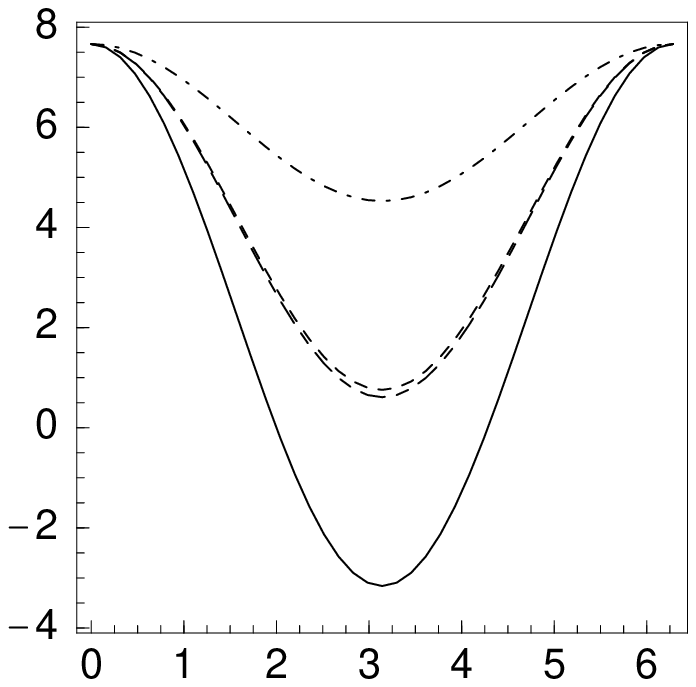,height=7.cm,width=7.cm}}}
\put(80,70){\makebox(0,0)[bl]{$10^{13} \cdot \Delta a_e$}}
\put(150,-4){\makebox(0,0)[br]{ 
       $\varphi_{M^2_{E,13}, M^2_{L,13}, A_{13}, A_{31}}$}}
\end{picture}
\end{center}
\caption{ a) $d_e$ and b) SUSY contributions to $a_e$ 
 as a function of $\varphi_{M^2_{E,13}}$ (full line),
$\varphi_{M^2_{L,13}}$ (dashed line), $\varphi_{A^l_{13}}$ (dashed dotted line)
and  $\varphi_{A_{31}}$ (long-short dashed line) for $\phimu=0$.}
\label{fig:13mixing}
\end{figure}

Now the flavour violating terms 
$M^2_{L,13} = 1500$~GeV$^2$, $M^2_{E,13} = 2000$~GeV$^2$,
$A_{31} =A_{13} = 20$~GeV are introduced.
This yields  
BR$(\tau \to e \, \gamma) = 1.05 \cdot 10^{-6}$. The effect of the
phases is shown in \fig{fig:13mixing}.
 Each individual contribution from the various phases
$\varphi_{M^2_{E,13}}$, $\varphi_{M^2_{L,13}}$, $\varphi_{A_{13}}$
and $\varphi_{A_{31}}$ is similar in size to that of $\phimu$.
If only one of these phases would generate the electron EDM, it would have
to be very near zero or $\pi$ because the effect is of order $10^{-23}$ ecm
as seen in the figure. But if there are several contributions, 
the phases can
be arbitrarily large, since various contributions can cancel each other.
Such a cancellation is not obvious, because the bounds on $\Delta a_e$ 
and BR$(\tau \to e \, \gamma)$ must be satisfied and the parameters are
already constrained.
The dependence of these quantities on the phases looks similar
to that of  BR$(\mu \to e \, \gamma)$ in Fig.~\ref{fig:12mixing} and 
are not shown here.

%
\begin{figure}
\setlength{\unitlength}{1mm}
\begin{center}
\begin{picture}(150,72)
\put(0,0){\mbox{\epsfig{figure=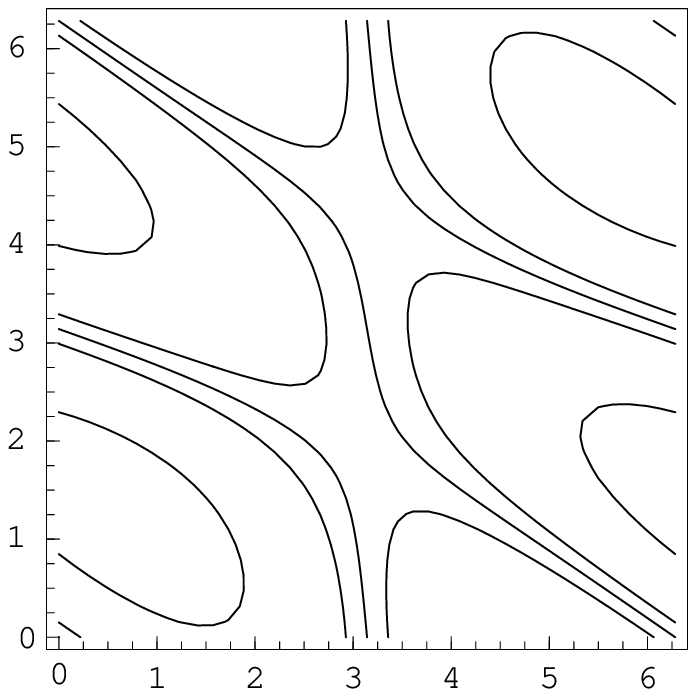,height=7cm,width=7cm}}}
\put(-5,73){\makebox(0,0)[bl]{{\bf a)}}}
\put(10,47){\makebox(0,0)[bl]{5}}
\put(56,53){\makebox(0,0)[bl]{5}}
\put(11,11){\makebox(0,0)[bl]{-5}}
\put(63,21){\makebox(0,0)[bl]{-5}}
\put(24,22){\makebox(0,0)[bl]{-1}}
\put(39,11){\makebox(0,0)[bl]{1}}
\put(24,35){\makebox(0,0)[bl]{1}}
\put(35,20){\makebox(0,0)[bl]{0}}
\put(36,43){\makebox(0,0)[bl]{0}}
\put(37,51){\makebox(0,0)[bl]{0}}
\put(42,36){\makebox(0,0)[bl]{-1}}
\put(40,65){\makebox(0,0)[bl]{1}}
\put(30,61){\makebox(0,0)[bl]{-1}}
\put(0,71){\makebox(0,0)[bl]{$\varphi_{M^2_{L,13}}$}}
\put(70,-4){\makebox(0,0)[br]{ $\phimu$}}
\put(77,0){\mbox{\epsfig{figure=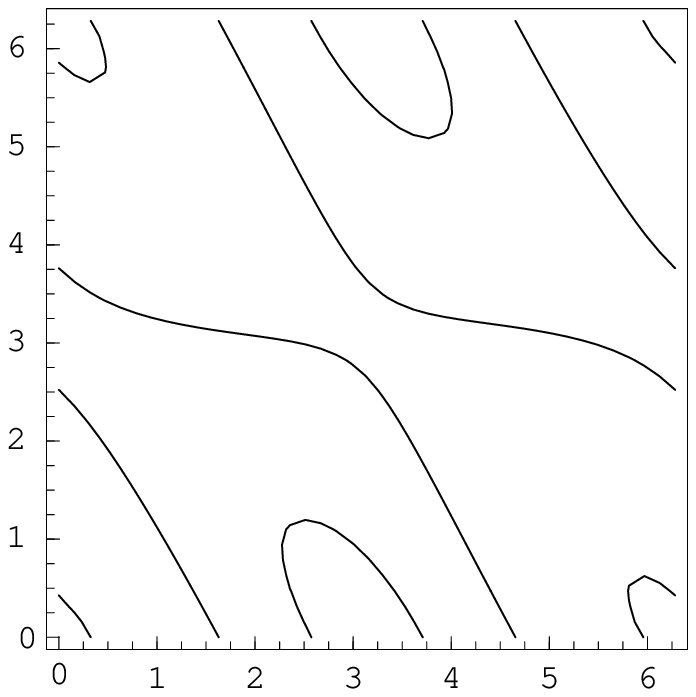,height=7.cm,width=7.cm}}}
\put(75,73){\makebox(0,0)[bl]{{\bf b)}}}
\put(110,12){\makebox(0,0)[bl]{1}}
\put(115,60){\makebox(0,0)[bl]{1}}
\put(85,19){\makebox(0,0)[bl]{5}}
\put(95,39){\makebox(0,0)[bl]{5}}
\put(125,39){\makebox(0,0)[bl]{5}}
\put(135,59){\makebox(0,0)[bl]{5}}
\put(83,10){\makebox(0,0)[bl]{10}}
\put(83,59){\makebox(0,0)[bl]{10}}
\put(141,13){\makebox(0,0)[bl]{10}}
\put(140,62){\makebox(0,0)[bl]{10}}
\put(80,71){\makebox(0,0)[bl]{$\varphi_{M^2_{L,13}}$}}
\put(147,-4){\makebox(0,0)[br]{ $\phimu$}}
\end{picture}
\end{center}
\caption{ a) $ 10^{24} d_e$ and b) $10^7$ BR$(\tau \to e \, \gamma)$ in
the $\phimu$ -- $\varphi_{M^2_{L,13}}$ plane.}
\label{fig:contourMuML13}
\end{figure}

In \fig{fig:contourMuML13} we show the contour plot for $d_e$ as a
function of $\varphi_\mu$ and $\varphi_{M^2_{L,13}}$. One can see that
a very small $d_e$ consistent with the experimental upper bound can be
obtained for all (!) values of $\phimu$ provided that also the phase
$\varphi_{M^2_{L,13}}$ of the flavour violating parameter $M^2_{L,13}$
is large. There are roughly two allowed regions. In one region, the two
phases are equal and opposite and there is a cancellation between the
lepton flavour conserving and the lepton flavour violating
contributions. In this case, the phase of $\mu$ can be large indeed.
In the other region, $\phimu$ is around $\pi$ and there is only a weak
dependence on $\varphi_{M^2_{L,13}}$. In this situation, the
contribution from ${M^2_{L,13}}$ is not important for the dipole
moment. Note that here only the phases shown
in the plot are varied, while the others are set equal to $0$.

As can be seen from \fig{fig:contourMuML13}b,  the decay
rate for $\tau \to e \, \gamma$  varies within an order of magnitude over
the plot. However, if further experiments would establish a considerably
lower limit or measure the branching ratio with 50\% or better, 
the phases could be severely constrained. This underlines
clearly the strength of a combined analysis, once the basic
supersymmetric parameters are known. We also note that the
'landscape' in the plots is quite dramatic and that  the allowed
regions are narrow. 

\begin{figure}
\setlength{\unitlength}{1mm}
\begin{center}
\begin{picture}(150,73)
\put(0,1){\mbox{\epsfig{figure=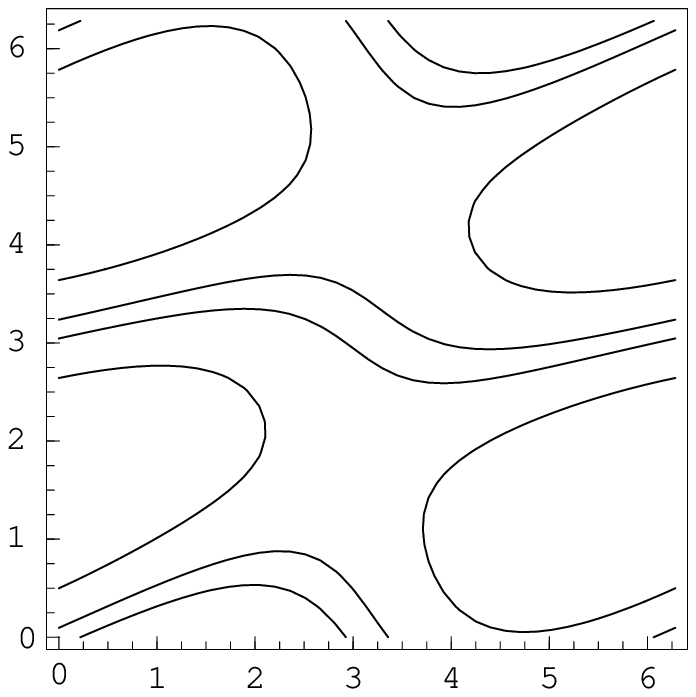,height=7cm,width=7cm}}}
\put(-5,73){\makebox(0,0)[bl]{{\bf a)}}}
\put(10,47){\makebox(0,0)[bl]{-5}}
\put(56,43){\makebox(0,0)[bl]{-5}}
\put(10,16){\makebox(0,0)[bl]{5}}
\put(56,9){\makebox(0,0)[bl]{5}}
\put(24,8){\makebox(0,0)[bl]{-1}}
\put(37,11){\makebox(0,0)[bl]{1}}
\put(55,67){\makebox(0,0)[bl]{1}}
\put(37,31){\makebox(0,0)[bl]{1}}
\put(37,36){\makebox(0,0)[bl]{-1}}
\put(37,61){\makebox(0,0)[bl]{-1}}
\put(1,72){\makebox(0,0)[bl]{$\varphi_{M^2_{E,13}}$}}
\put(70,-4){\makebox(0,0)[br]{ $\phimu$}}
\put(77,0){\mbox{\epsfig{figure=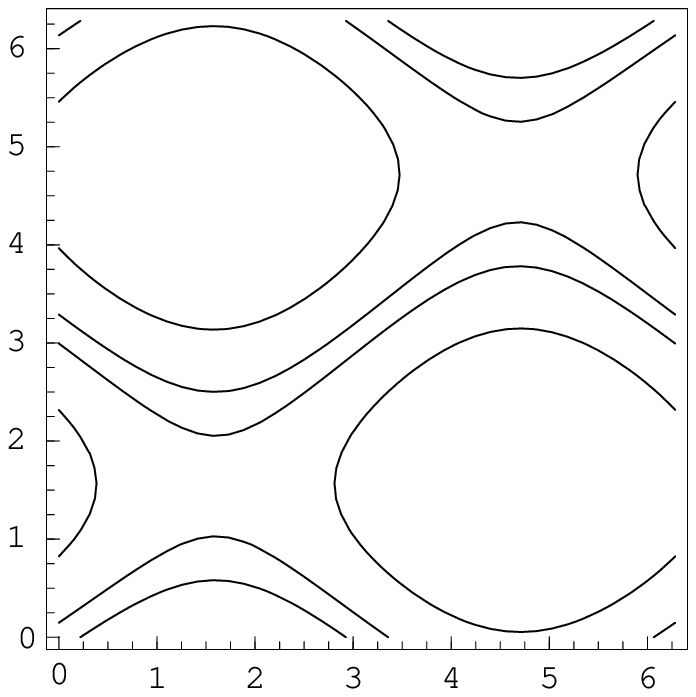,height=7.cm,width=7.cm}}}
\put(75,73){\makebox(0,0)[bl]{{\bf b)}}}
\put(90,63){\makebox(0,0)[bl]{-5}}
\put(138,55){\makebox(0,0)[bl]{-5}}
\put(136,9){\makebox(0,0)[bl]{5}}
\put(86,19){\makebox(0,0)[bl]{5}}
\put(98,8){\makebox(0,0)[bl]{-1}}
\put(98,17){\makebox(0,0)[bl]{1}}
\put(98,23){\makebox(0,0)[bl]{1}}
\put(98,31){\makebox(0,0)[bl]{-1}}
\put(133,57){\makebox(0,0)[bl]{-1}}
\put(133,65){\makebox(0,0)[bl]{1}}
\put(80,72){\makebox(0,0)[bl]{$\varphi_{A_{31}}$}}
\put(147,-4){\makebox(0,0)[br]{ $\phimu$}}
\end{picture}
\end{center}
\caption{ $ 10^{24} d_e$, a) in the $\phimu$ -- $\varphi_{M^2_{E,13}}$ plane
b)  in the $\phimu$ -- $\varphi_{A_{31}}$ plane.}
\label{fig:contourMuME13}
\end{figure}

In Fig.~\ref{fig:contourMuME13} the situation for the phases of
the flavour violating parameters
$M^2_{E,13}$ and $A_{13}$ are shown. These terms have a strong
interplay with the phase of $\mu$ as seen by the 'dramatic' landscape.
In both cases, the phase is essentially limited to  zero and
$\pi$ (recall, phases not shown in the plot 
are set equal to zero). On the other hand, the phase of $\mu$ is
not restricted.




\subsection{$\tilde \mu$-$\tilde \tau$  mixing}

\begin{figure}
\setlength{\unitlength}{1mm}
\begin{center}
\begin{picture}(150,70)
\put(0,0){\mbox{\epsfig{figure=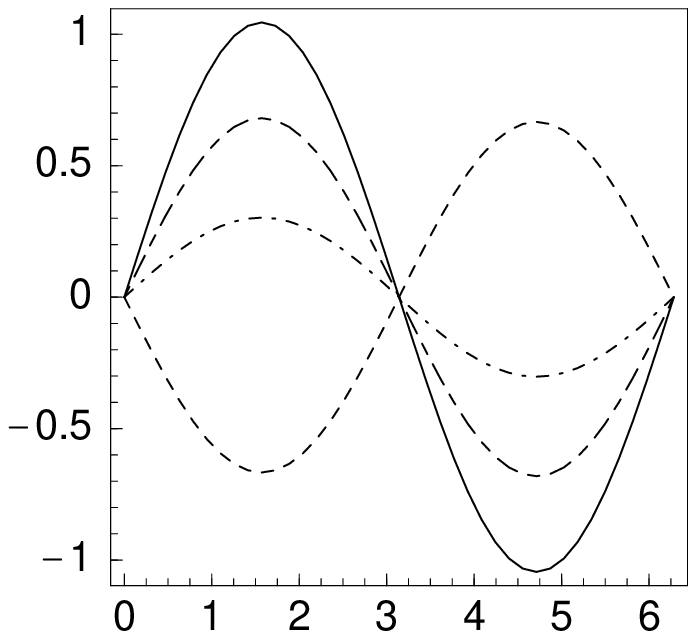,height=7cm,width=7cm}}}
\put(-5,71){\makebox(0,0)[bl]{{\bf a)}}}
\put(0,70){\makebox(0,0)[bl]{$10^{23} \cdot d_\mu$}}
\put(70,-2){\makebox(0,0)[br]{
       $\varphi_{M^2_{E,23}, M^2_{L,23}, A_{23}, A_{32}}$}}
\put(77,0){\mbox{\epsfig{figure=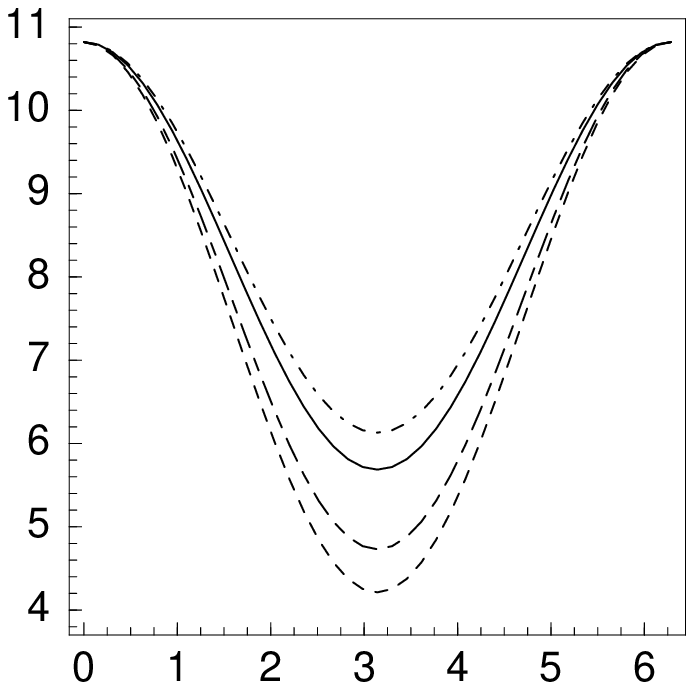,height=7.cm,width=7.cm}}}
\put(75,71){\makebox(0,0)[bl]{{\bf b)}}}
\put(80,70){\makebox(0,0)[bl]{$10^7 \cdot$ BR$(\tau \to \mu \, \gamma)$}}
\put(150,-2){\makebox(0,0)[br]{
       $\varphi_{M^2_{E,23}, M^2_{L,23}, A_{23}, A_{32}}$}}
\end{picture}
\end{center}
\caption{ a) $d_\mu$ and b)  BR$(\tau \to \mu \, \gamma)$
 as a function of $\varphi_{M^2_{E,23}}$ (full line),
$\varphi_{M^2_{L,23}}$ (dashed line), $\varphi_{A_{23}}$ (dashed dotted line)
and  $\varphi_{A_{32}}$ (long-short dashed line)  for $\phimu=0$..}
\label{fig:23mixing}
\end{figure}

In this subsection we consider mixing between the second and the third
generation by adding the following flavour violating terms to our
reference point: $M^2_{L,23} = 1500$~GeV$^2$, $M^2_{E,23} =
2000$~GeV$^2$, $A_{32} =A_{23} = 20$~GeV, yielding BR$(\tau \to \mu \,
\gamma) = 1.0 \cdot 10^{-6}$. The effect of the phases is shown in
\fig{fig:23mixing}. We see the branching ratio BR$(\tau \to \mu \,
\gamma)$ can vary by a factor two when changing the phases from 0 to
$\pi$.
If one takes $\varphi_\mu=\pi/2$ then one finds 
$d_\mu=(-2.7 \pm 0.1) 10^{-22}$ecm varying only slightly with the flavour 
violating phases.  Note that this value is 
clearly in the reach of future experiments \cite{edipol}. 
In the case of BR$(\tau \to \mu \, \gamma)$ we find
a range of $2.5 \cdot 10^{-7}$--$8.5 \cdot 10^{-7}$ and the functional
dependence on the phases are shifted by $\pi/2$ for $\varphi_{M^2_{E,23}}$
and $-\pi/2$ in the remaining cases.

\subsection{The three generation case}

Finally, we allow for the most general mixing in the slepton sector.
We take our reference point and add all possible phases and generation
mixing terms. The moduli of the off-diagonal terms are between zero and
the following upper bounds: $|M^2_{E,12}|, |M^2_{L,12}| \le 10$~GeV$^2$,
$|M^2_{E,13}|, |M^2_{E,23}|, |M^2_{L,13}|, |M^2_{L,23}| \le 1000$~GeV$^2$,
$|A_{12}|, |A_{21}| \le 0.05$~GeV, 
$|A_{13}|, |A_{31}|, |A_{23}|, |A_{32}| \le 20$~GeV. All phases
are varied in the range between 0 and 2 $\pi$. 

\begin{figure}
\setlength{\unitlength}{1mm}
\begin{center}
\begin{picture}(150,70)
\put(0,0){\mbox{\epsfig{figure=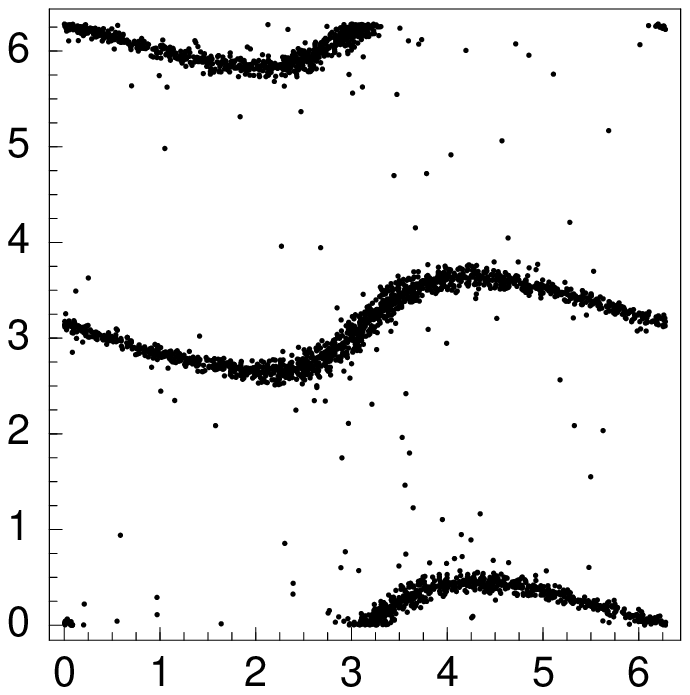,height=7.cm,width=7.cm}}}
\put(-5,71){\makebox(0,0)[bl]{{\bf a)}}}
\put(0,70){\makebox(0,0)[bl]{$\phimu$}}
\put(70,-3){\makebox(0,0)[br]{$\phim$}}
\put(80,0){\mbox{\epsfig{figure=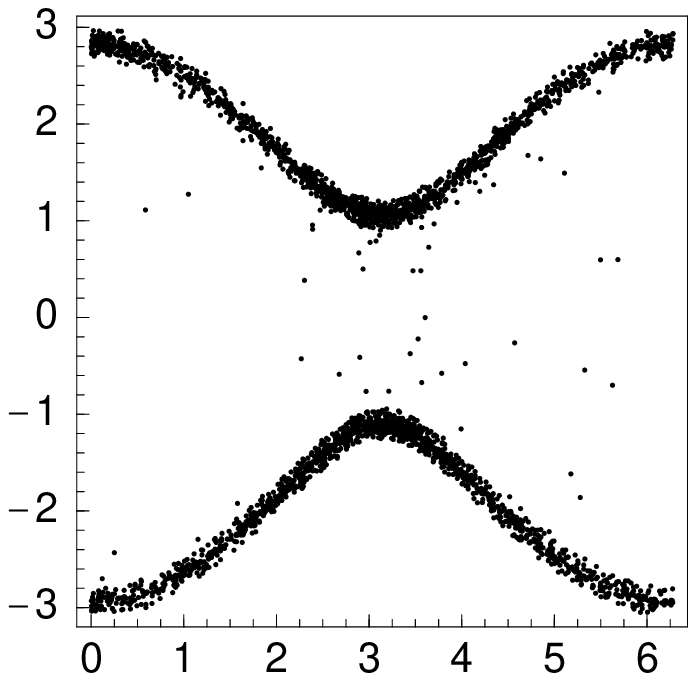,height=7.cm,width=7.cm}}}
\put(75,71){\makebox(0,0)[bl]{{\bf b)}}}
\put(80,70){\makebox(0,0)[bl]{$10^9 \cdot \Delta a_\mu$}}
\put(150,-3){\makebox(0,0)[br]{$\phim$}}
\end{picture}
\end{center}
\caption{ a) Allowed regions in the $\phimu$--$\phim$; b) SUSY contribution
$\Delta a_\mu$ to the anomalous magnetic moment of the muon. Here we
have taken the most general form for the slepton mixings}
\label{fig:MuM1plane3gen}
\end{figure}

\fig{fig:MuM1plane3gen} is a scatter plot of the allowed values of  $\phimu$
and $\phim$ obeying all constraints from the EDMs and the rare lepton
decays. Comparing \fig{fig:MuM1plane3gen} with \fig{fig:MuM1plane} one sees
that maximal values for both $\phimu$
and $\phim$ are possible. This is again due to cancellations between
the lepton flavour conserving and the lepton flavour violating contributions.
Note that such cancellations are possible even for slepton masses as small
as 200 GeV. 
In \fig{fig:MuM1plane3gen}b we show the SUSY contribution $\Delta a_\mu$
to the anomalous magnetic moment of the muon as a function of $\phim$,
varying all parameters and phases in the range given above and fulfilling
the constraints from the EDMs and the rare lepton decays. 

\begin{figure}
\setlength{\unitlength}{1mm}
\begin{center}
\begin{picture}(150,70)
\put(0,0){\mbox{\epsfig{figure=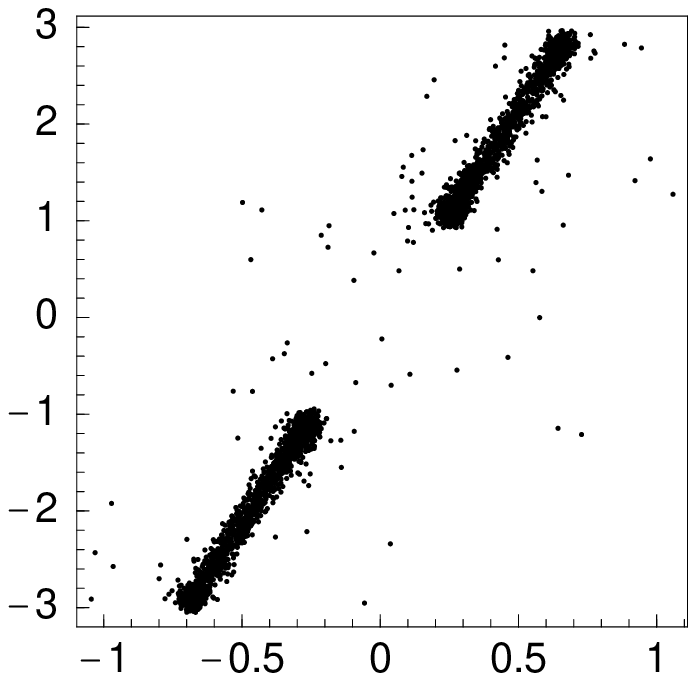,height=7.cm,width=7.cm}}}
\put(-5,71){\makebox(0,0)[bl]{{\bf a)}}}
\put(0,70){\makebox(0,0)[bl]{$10^9 \cdot \Delta a_\mu$}}
\put(70,-4){\makebox(0,0)[br]{$10^{13} \cdot \Delta a_e$}}
\put(80,0){\mbox{\epsfig{figure=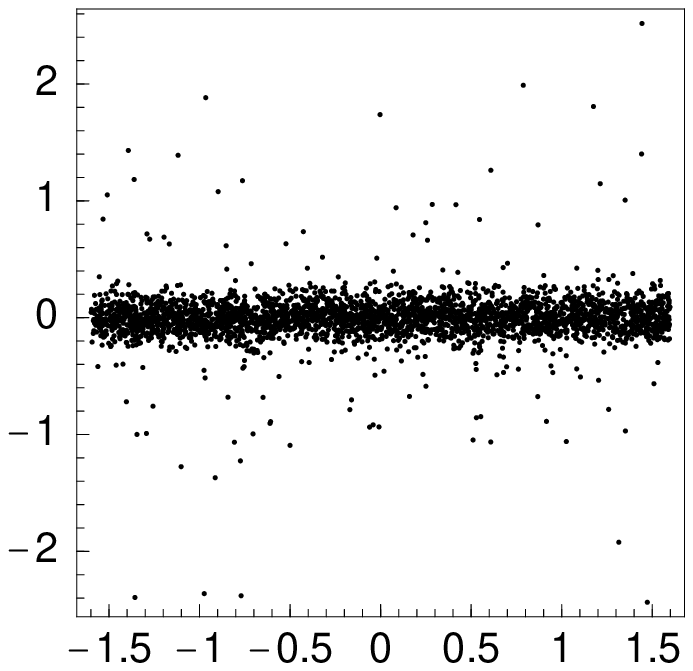,height=7.cm,width=7.cm}}}
\put(75,71){\makebox(0,0)[bl]{{\bf b)}}}
\put(80,70){\makebox(0,0)[bl]{$10^{22} \cdot d_\mu$}}
\put(150,-4){\makebox(0,0)[br]{$10^{27} \cdot d_e$}}
\end{picture}
\end{center}
\caption{ Correlation between a) $\Delta a_e$ and $\Delta a_\mu$
 and b) $d_e$ and $d_\mu$ .}
\label{fig:AeAmuplane}
\end{figure}

As pointed out in \cite{feng}, LFV leads to the violation of the naive
scaling relations like $d_e/d_\mu \simeq m_e/m_\mu$. Similarly one
also expects deviations from the relation $\Delta a_e/\Delta a_\mu
\propto (m_e/m_\mu)^2$, in particular as a consequence of the phases
of the parameters. \fig{fig:AeAmuplane} shows our results for
$\Delta a_e$ versus $\Delta a_\mu$ and $d_e$ versus $d_\mu$. 
 One sees that the naive relation
$\Delta a_e /\Delta a_\mu \propto (m_e/m_\mu)^2$ is largely 
maintained after imposing the experimental constraints arising from
EDMs and rare decays even if one allows for the most general flavour
structure. However, there are parameter points where the simple
$\Delta a_\mu$ -- $\Delta a_e$ scaling is violated, as has also been noted
by the authors of Refs.~\cite{feng,strumia}. Of interest is the 'hole' 
in \fig{fig:AeAmuplane}a which excludes vanishing corrections.

The situation is completely
different in the case of the electric dipole moments where the correlation
between $d_e$ and $d_\mu$ is completely destroyed once all
possible flavour violating parameters are taken into account
\footnote{Other possibilities of violations of the scaling relations
have been presented in \cite{Babu:2000dq}}. The
reason for the difference between EDMs and the MDMs is that in the case
of the $d_e$ cancellations of at least of two orders of magnitude are required
to satisfy the experimental bounds implying that $d_e$ is no longer
 proportional  to $m_e$. We have checked that in the case,
where a larger modulus of $d_e$ is allowed, the proportionality to $m_e$
is restored except for the region around 0.
In addition we have checked that the ratio 
$d_\mu/d_\tau$ is still proportional to $m_\mu/m_\tau$.


\begin{figure}
\setlength{\unitlength}{1mm}
\begin{center}
\begin{picture}(150,155)
\put(0,85){\mbox{\epsfig{figure=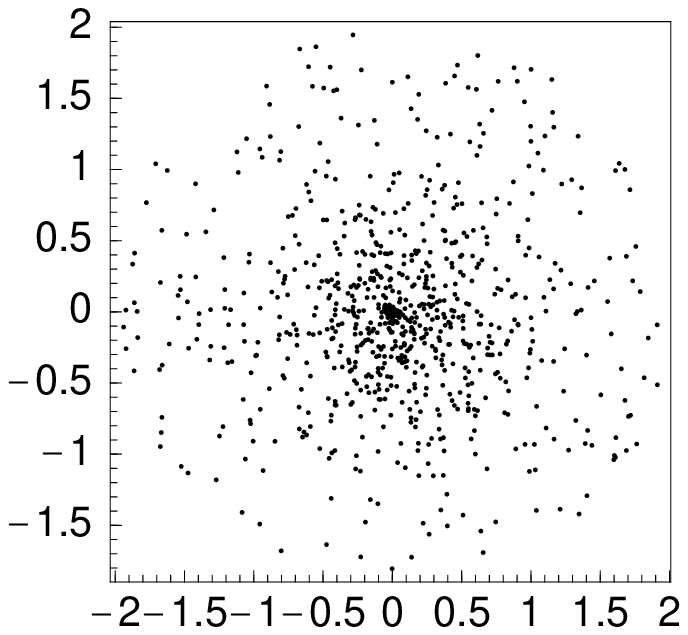,height=7.cm,width=7.cm}}}
\put(-5,155){\makebox(0,0)[bl]{{\bf a)}}}
\put(0,154){\makebox(0,0)[bl]{Im($M^2_{E,13}$)~[$10^3$ GeV$^2$]}}
\put(70,82){\makebox(0,0)[br]{Re($M^2_{E,13}$)~[$10^3$ GeV$^2$]}}
\put(80,85){\mbox{\epsfig{
                 figure=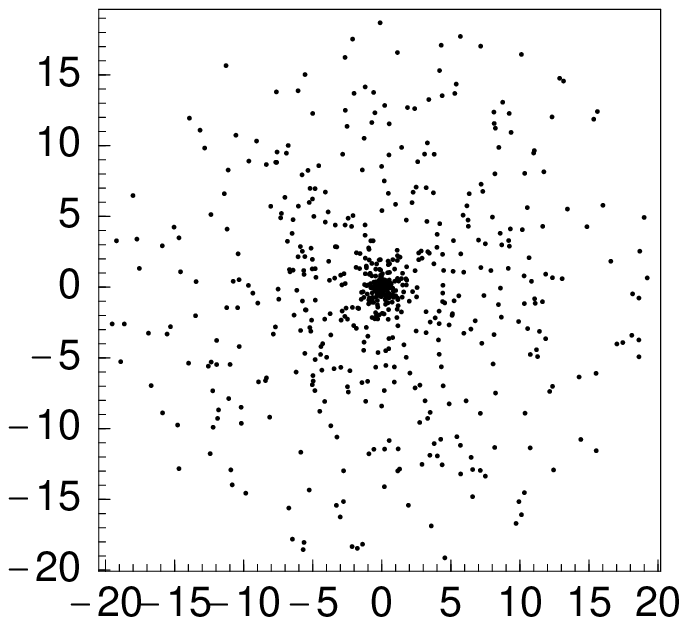,height=7.cm,width=7.cm}}}
\put(75,155){\makebox(0,0)[bl]{{\bf b)}}}
\put(80,154){\makebox(0,0)[bl]{Im($A_{13}$)~[GeV]}}
\put(150,82){\makebox(0,0)[br]{Re($A_{13}$)~[GeV]}}
\put(0,0){\mbox{\epsfig{figure=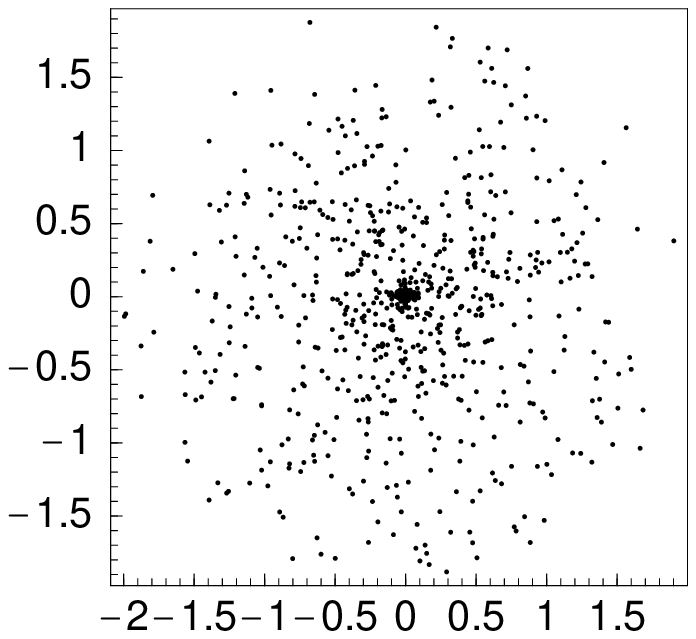,height=7cm,width=7cm}}}
\put(-5,70){\makebox(0,0)[bl]{{\bf c)}}}
\put(0,69){\makebox(0,0)[bl]{Im($M^2_{L,13}$)~[$10^3$ GeV$^2$]}}
\put(70,-4){\makebox(0,0)[br]{Re($M^2_{L,13}$)~[$10^3$ GeV$^2$]}}
\put(80,0){\mbox{\epsfig{
                 figure=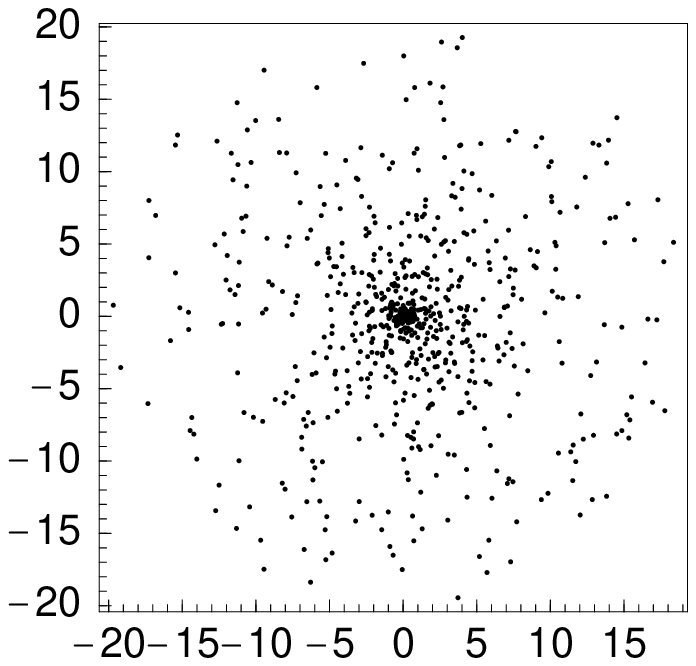,height=7.0cm,width=7.0cm}}}
\put(75,70){\makebox(0,0)[bl]{d)}}
\put(80,69){\makebox(0,0)[bl]{Im($A_{31}$)~[GeV]}}
\put(150,-4){\makebox(0,0)[br]{Re($A_{31}$)~[GeV]}}
\end{picture}
\end{center}
\caption{ Real and imaginary parts of $M^2_{E,13}$, 
  $M^2_{L,13}$, $A_{13}$, and $A_{31}$  allowed by the experimental
 constraints. We have taken $\phimu=\pi/2$ and
 the remaining phases have been varied as described in the the text.}
\label{fig:ReImME}
\end{figure}

\fig{fig:ReImME} shows the allowed regions for the complex
parameters  $M^2_{E,13}$, $M^2_{L,13}$, $A_{13}$ and $A_{31}$
respectively, for $\phimu=\pi/2$. All other phases
have been varied in the range $(0,2 \pi)$.
Again, the allowed regions are large. Note, that the
$|M^2_{E,13}|$, $|M^2_{L,13}|$ can go up to 5\% of $|M^2_{E,33}|$
and $|M^2_{L,33}|$, respectively. $|A_{13}|$ and $|A_{31}|$ can have the
same order of magnitudes as $|A_{33}|$. 
In case of $\phimu=0$ roughly the same areas would be allowed. 
The major differences compared to $\phimu=\pi/2$ are: 
(i) The moduli of the $A$ parameters
are smaller by about 25\%. (ii) There are less points with large moduli.

\section{Conclusions}

In this paper we have studied the most general mass matrices for
sleptons within the MSSM, including left--right mixing, flavour mixing
and complex phases. 
In particular, we have analysed the implications on the
phases coming from the experimental restrictions on anomalous magnetic and
electric dipole moments of the charged leptons and on the rare decays
$\mu \to e \gamma$, $\tau \to e \gamma$ and $\tau \to \mu \gamma$.
For the basic SUSY parameters such as
the masses of the super-particles we have used the values discussed
in the Snowmass report \cite{Allanach:2002nj}.

Since there are many free parameters in a general
scenario, we have first considered several special situations:

(i) If flavour violation in the slepton sector is
negligible (no off-diagonal matrix elements)
we recover the known results for the phases: In general, all 
possible phases, especially the phase $\varphi_\mu$ of the $\mu$ parameter 
must
be small (or $\pi$) in order to be consistent  with the electric
dipole moments. Only in the case that the phases of $M_1$ and $A_{11}$
are correlated to the phase of $\mu$, the phases of $M_1$ and $A_{11}$
can be maximal due to cancellations of different contributions to $d_e$.

(ii) In case there is flavour mixing only in the selectron--smuon sector,
the moduli of the flavour
violating parameters in the mass matrices have to be small to satisfy
the experimental bound on $\mu \to e \gamma$. Therefore, the
effects of their phases is small even if they are maximal.

(iii) If there is only mixing between  selectrons and 
staus, each individual contribution to $d_e$
due to the phases $\phimu$, $\phim$, $\varphi_{M^2_{E,13}}$, 
$\varphi_{M^2_{L,13}}$,
$\varphi_{A_{13}}$, and $\varphi_{A_{31}}$  is of similar size.
If only one of these phases is non-vanishing, it must be small if the 
slepton masses are O(100)~GeV. However, if two or more phases are present, 
all of them including $\phimu$
could be large because the various contributions to $d_e$ may
cancel each other.

(iv) And finally, if only smuons and staus mix, the present
experiments do not limit the phases of the
mixing parameters. However, the planned
new measurement of the muon EDM being sensitive to 
$d_\mu \simeq 10^{-24}$~ecm \cite{edipol} could give restrictions 
on phases if only one is present, otherwise combinations of various phases
will be constrained.

In the general case with arbitrary three--generation mixing, cancellations
between various LFC and LFV contributions to $d_e$ are easily possible.
The numerical analysis has lead to two main results:

(i) Despite the large number of unknown parameters, significant restrictions
on the allowed ranges  are obtained. A good example is given in 
\fig{fig:MuM1plane3gen} 
From
\fig{fig:MuM1plane3gen} we see that the allowed range for the new 
contributions to $g-2$ are limited by the two wiggly bands. So if
the theoretical analysis of $ g - 2 $ in the standard model
would yield that $\Delta a$ is in the range $(1-2) \cdot 10^{-9}$, the
phase $\phi_{U(1)}$ would have to be near $90$ or $270$ degrees.
Similarly  \fig{fig:ReImME} a and c  limit the values of
$M^2_{E,13}$ and $M^2_{L,13}$ to about $2 \cdot 10^3 GeV^2$,
about $2$--$3$ \% of the values of the diagonal elements; in contrast,
$A_{13}$ and $A_{31}$ are much less constrained.

\fig{fig:AeAmuplane} a) shows an unexpected hole in the allowed ranges of 
the muon and electron $g-2$, and \fig{fig:AeAmuplane} b) the
surprising independence of the respective electric dipole moments.
This is a  clear indication \cite{feng} that the presence of lepton
flavour violating phases leads to large deviations of the scaling relations
such as $d_e/d_\mu \simeq m_e/m_\mu$, but to much smaller modifications
for the scalings of $\Delta a_e/\Delta a_l$ etc. In the case
of the electron, the mass is so small that other contribution 
are also important and may swamp out the mass dependence
almost completely \footnote{If the range in \fig{fig:AeAmuplane} is
increased, a weak mass dependence is visible.}. Therefore, it is  
possible that the EDM's of $\mu$ and $\tau$ are larger than expected 
from "naive" scaling.

(ii) The lepton flavour violating parameters as well as $\mu$,
$M_1$ and $A_{ii}$ (i=1,2,3) can have large phases, despite the 
stringent limits on $CP$ violation.
In particular, the phase of
the parameter $\mu$ can be maximal even for O(100)~GeV slepton masses,
in contrast to naive expectations.
Therefore, the phases of the
supersymmetric parameters can be as large as those in the standard
model and need not be artificially small.  While we have used one
of the Snowmass points for our presentation of detailed numerical results, 
we have checked that the qualitative
features of our results do not depend on this specific choice.

\section*{Acknowledgments}

We thank W.~Bernreuther for useful discussions. 
This work was supported by the European Community's Human
Potential Programme under contract HPRN-CT-2000-00149, by
the `Fonds zur F\"orderung der wissenschaftlichen Forschung' of Austria,
projects No.~P13139-PHY and No.~P16592-N02
 and by the Swiss 'Nationalfonds'. W.P.~has been
supported by the Erwin
Schr\"odinger fellowship No. J2272 of the `Fonds zur
F\"orderung der wissenschaftlichen Forschung' of Austria.

\begin{appendix}

\section{Slepton couplings}

Here we collect the formulas for the various slepton couplings in the
most general form.
The couplings $l_i$-$\tilde \chi^+_j$-$\tilde \nu_k$ are given by
\begin{eqnarray}
d^L_{ijk} &=&  - g V_{j1} \sum_{r=1}^3 (R^{\tilde \nu}_{kr})^* R^l_{L,ir} \\
d^R_{ijk} &=&  U_{j2} \sum_{r,s=1}^3 (R^{\tilde \nu}_{kr})^* Y^E_{rs}
               R^l_{R,is}
\end{eqnarray}
where $U$ and $V$ are the chargino mixing matrices, $R^{\tilde \nu}$ is
the sneutrino mixing matrix, $R^l_L$ and $R^l_R$ are the left and right 
mixing matrix of the left and right charged leptons, respectively. Note that
in the absence of right handed neutrinos the latter two can be chosen to
be the unit matrix without loss of generality.
The couplings $l_i$-$\tilde \chi^0_j$-$\tilde l_k$ are given by
\begin{eqnarray}
c^L_{ijk} &=&  f^R_j \sum_{r=1}^3 (R^{\tilde l}_{k,3+r})^*  R^l_{R,ir}
        - N^*_{j3} \sum_{r,s=1}^3 (R^{\tilde l}_{k,r})^* Y^E_{rs} R^l_{R,is} \\
c^R_{ijk} &=& f^L_j   \sum_{r=1}^3 (R^{\tilde l}_{k,r})^*  R^l_{L,ir}
        - N_{j3} \sum_{r,s=1}^3 (R^{\tilde l}_{k,3+r})^* (Y^E_{sr})^*
                               R^l_{L,is} \\
f^R_j &=& - \sqrt{2} g' N_{j1} \\
f^L_j &=& \frac{1}{\sqrt{2}} \left( g' N_{j1} - g N_{j2} \right)
\end{eqnarray}
where $N$ is the neutralino mixing matrix and $R^{\tilde l}$ is the mixing
matrix of the charged sleptons.

\section{Loop functions}

For completeness  we display here the loop functions used:
\begin{eqnarray}
F_1(x) &=& - \frac{2 + 3 x - 6 x^2 + 3 x^3 + 6 x \log x}{6 (1-x)^4} \\
F_2(x) &=& \frac{1 - 6 x + 3 x^2 + 2 x^3 - 6 x^2 \log x}{6 (1-x)^4} \\
F_3(x) &=& - \frac{1 - x^2 + 2 x \log x}{(1-x)^3}  \\
F_4(x) &=& \frac{1 - 4 x + 3 x^2 - 2 x^2 \log x}{ (1-x)^3}
\end{eqnarray}

\end{appendix}

\end{document}